\documentclass[onecolumn]{aa}
\usepackage[utf8]{inputenc} 
\usepackage{multirow}
\usepackage{lscape}

\usepackage{graphicx}
\usepackage{txfonts}
\usepackage{hyperref}
\hypersetup{
    colorlinks=true,
    linkcolor=red,
    filecolor=magenta,      
    urlcolor=magenta,
    citecolor=blue,
}

\begin{document}

   \title{Insights into Chromospheric Large-Scale Flows using Nobeyama 17 GHz Radio Observations}

   \subtitle{I. The Differential Rotation Profile}

   \author{Srinjana Routh
          \inst{1,2}
          \and
          Anshu Kumari
          \inst{3,4} 
          \fnmsep\thanks{Corresponding Author \email{anshu@prl.res.in}}
                    \and
                              Vaibhav Pant \inst{1}
        \and
         Jaydeep Kandekar\inst{5}
                   \and
        Dipankar Banerjee \inst{6,7,8}
                    \and
        Mohd. Saleem Khan \inst{2}
                    \and
        Dibya Kirti Mishra \inst{1,2}
          }

   \institute{Aryabhatta Research Institute of Observational Sciences, Nainital-263002, Uttarakhand, India
         \and
         Department of Applied Physics, Mahatma Jyotiba Phule Rohilkhand University, Bareilly-243006, Uttar Pradesh, India 
            \and
  Department of Physics, University of Helsinki, P.O. Box 64, FI-00014, Helsinki, Finland
         \and
        Udaipur Solar Observatory, Physical Research Laboratory, Dewali, Badi Road, Udaipur-313 001, Rajasthan, India 
   \and
       Department of Physics, Ahmednagar College, Station Road, Ahilyanagar-414001, Maharashtra, India
       \and
       Indian Institute of Space Science and technology, Valiamala, Thiruvananthapuram - 695 547,Kerala, India
       \and
       Indian Institute of Astrophysics, Koramangala, Bangalore 560034, India
       \and
       Center of Excellence in Space Sciences India, IISER Kolkata, Mohanpur 741246, West Bengal, India
             }


 
  \abstract
   {Although the differential rotation rate on the solar surface has long been studied using optical and extreme ultraviolet (EUV) observations, associating these measurements to specific atmospheric heights remains challenging due to the temperature-dependent emission of tracers observed in EUV wavelengths. Radio observations, being primarily influenced by coherent plasma processes and/or thermal bremsstrahlung, offer a more height-stable diagnostic and thus provide an independent means to test and validate rotational trends observed at other EUV wavelengths.}
   {We aim to characterize the differential rotation profile of the upper chromosphere using cleaned solar full-disc 17 GHz radio imaging from the Nobeyama Radioheliograph (NoRH), spanning a little over two solar cycles (1992–2020).}
   {A tracer-independent method based on automated image correlation was employed on daily full-disc 17 GHz radio maps. This method determines the angular velocities in 16 latitudinal bins of $15^{\circ}$ each by maximizing the 2D cross-correlation of overlapping image segments.}
   {The best-fit parameters for the differential rotation profile are A = 14.520 ± 0.006°/day, B = -1.443 ± 0.099°/day, and C = -0.433 ± 0.267°/day. These results suggest that the upper chromosphere rotates significantly faster than the photosphere at all latitudes, with a relatively flatter latitudinal profile. A very weak anti-correlation ($\rho_{s}=-0.383$ (94.73\%)) between the equatorial rotation rate and solar activity is also observed.}
   {Our findings reaffirm the potential of radio observations to probe the dynamics of the solar chromosphere with reduced height ambiguity. The overlap of the equatorial rotation rate ($A$) found in this study with that for $304$ {\AA} in the EUV regime lends additional support to the view that the equatorial rotation rates increase with height above the photosphere. Future coordinated studies at wavelengths with better-constrained height formation will be crucial for further understanding the complex dynamics of the solar atmosphere.}


 \maketitle
%

\section{Introduction}
Large-scale flows on the Sun play a pivotal role in shaping the global dynamics of the solar atmosphere and influencing the distribution of magnetic fields across various layers of the solar interior and exterior. Among these, differential rotation, empirically formulated by Eq. (\ref{diffequation}), is one of the most fundamental and extensively studied global motions, observed not only at the solar surface but across the full extent of the solar atmosphere. 
\begin{equation}
\Omega= A + B\sin^2{\theta} + C \sin^4{\theta}, 
\label{diffequation}
\end{equation}
The differential rotation of the solar photosphere has been widely studied through techniques including feature tracking, flux modulation, and other methods \citep[e.g.,][]{Newton1951,Snodgrass1983}. Furthermore, recent advancements in helioseismology have significantly enhanced our understanding of the latitudinal and radial rotation profiles within the solar interior \citep{Antia1998,Antia2008}. With growing observational capabilities spanning optical, Ultraviolet (UV), Extreme Ultraviolet (EUV), and radio wavelengths, the opportunity has emerged to reframe the problem in terms of a broader class of large-scale flows that pervade the atmosphere. From this perspective, differential rotation is not merely a parameterization of latitudinal shear, but a tracer of how angular momentum transport, plasma motions, and magnetic fields are coupled across spatial and temporal scales in the solar atmosphere.

However, despite centuries of investigation, the nature of how such flows evolve with height and temperature, and their interconnection across stratified atmospheric layers, remains incompletely understood due to the height ambiguity of tracers involved in the extraction of the differential rotation profile. Recent studies, particularly using the Atmospheric Imaging Assembly (AIA) onboard the Solar Dynamic Observatory \citep[SDO; e.g.,][]{Sharma2020,Routh2024}, have demonstrated that a possible connection between the rotation rate of the solar atmosphere might exist with height above the photosphere. These trends suggest that the atmospheric layers are not passively rotating but may be dynamically linked to deeper processes, possibly governed by the magnetic field topology. This line of reasoning has found support in works that compare atmospheric rotation with internal rotation profiles derived from helioseismology \citep[e.g.,][]{Badalyan2010,FinelyBrun2023}, hinting at connections mediated by magnetically anchored structures.

Despite such broad efforts, the phenomenon of differential rotation has been difficult to grasp in the upper and hotter solar atmosphere due to various reasons, including but not limited to the fast-changing nature of higher atmospheric tracers as well as the height ascribed to them. Although the faster rotation of the hotter solar atmosphere, specifically the chromosphere, has long been suggested through different methods and different datasets \citep{Livingston1969,Mishra2024,Routh2024}, an ambiguity in the exact height of the origin of the emission dataset in such studies has always left a gap to be fulfilled. Extending this inquiry into radio wavelengths offers an even less ambiguous insight into the chromosphere, transition region and corona. Radio observations, such as those from the Nobeyama Radioheliograph (NoRH), provide height-stable diagnostics of the upper chromosphere and transition region\citep[e.g., 34 and 17 GHz emission (8.8 and 17.6 mm, respectively); ][]{Selhorst2008}, thereby avoiding the temperature-based ambiguities that often affect EUV diagnostics. In an attempt to address this gap, this study utilizes a tracer-independent method to ascertain the spatial and temporal variation in the differential rotation of the solar atmosphere at well-defined heights above the photosphere as determined from radio emission. The article is arranged as follows: \autoref{section:sec2} discusses the observations and data analysis method, section \autoref{section:sec3} discusses the results.  Finally, \autoref{section:sec4} summarizes the present work and discusses the conclusion.

\section{Observation and data analysis}
\label{section:sec2}
\subsection{The Nobeyama Radioheliograph}
The Nobeyama Radioheliograph, operational since 1992 July till 2020 March, has been observing the Solar chromosphere and the transition region at frequencies of 34 and 17 GHz, respectively at a resolution of 10" \citep{NobeyamaInstrument1994}. The cleaned solar full-disc images \footnote{\url{https://solar.nro.nao.ac.jp/norh/images/daily/}} that are used in this study have dimensions of $512 \times512$ pixel\textsuperscript{2} with a pixel scale of $4.91''$ . These images are image registered and aligned such that the solar north coincides with the image north.

\begin{figure}
    \centering
    \includegraphics[width=\linewidth]{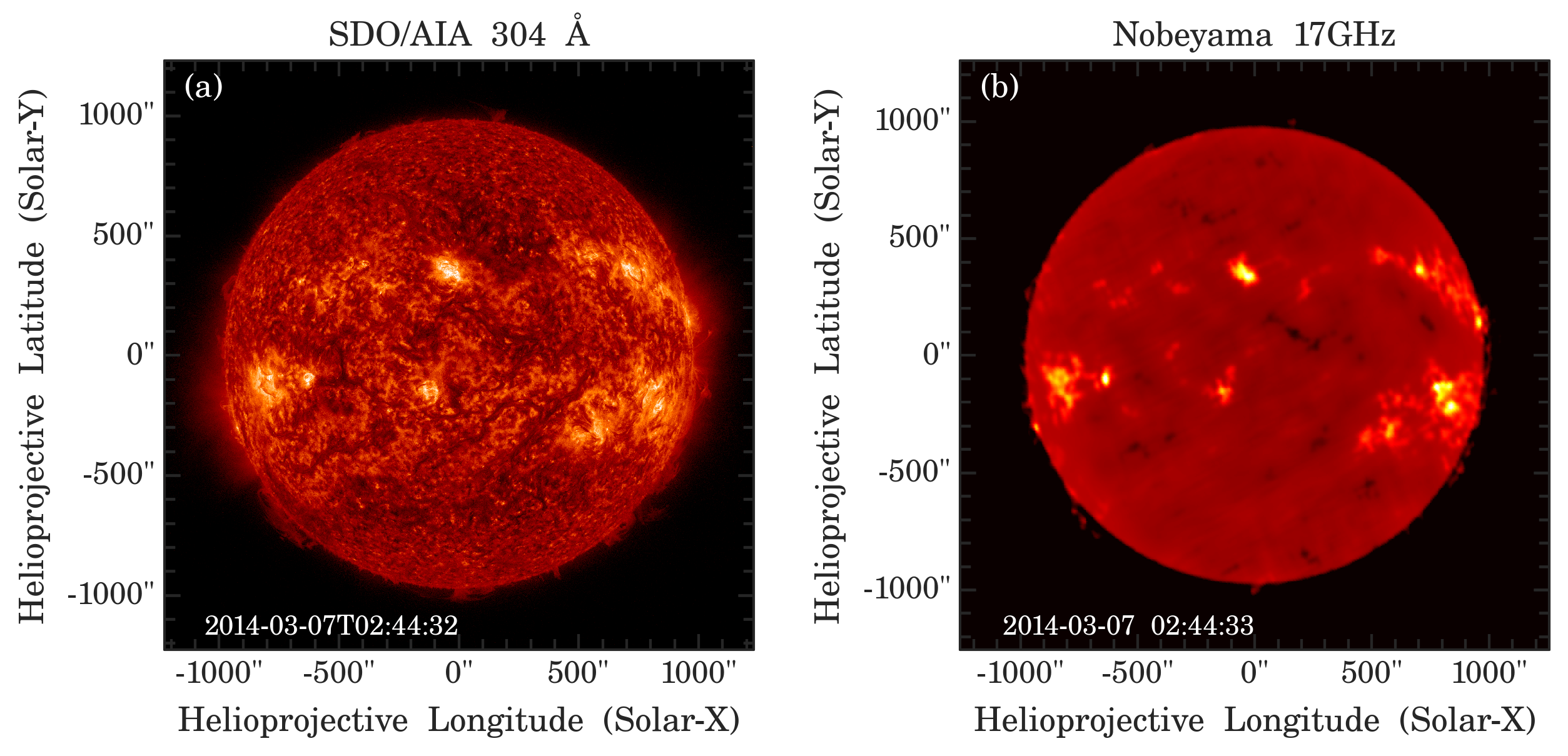}
    \caption{The Sun from (a) a space-based perspective as seen by the Atmospheric Imaging Assembly (AIA) atop the Solar Dynamic Observatory and (b) a ground-based perspective as seen by the Nobeyama Radioheliograph.}
    \label{fig:radio-aia}
\end{figure}

\subsection{Data pre-processing and method of image correlation}
Since the cleaned solar full-disc fits image data from the Nobeyama Radioheliograph primarily represents the brightness temperature of various features in the solar atmosphere, the maximum brightness temperature values in the dataset can vary significantly depending on the observed solar activities \citep{2003A&A...401.1143S}. This might create an issue when the method of image correlation, which primarily utilizes the intensity gradient information to minimize the brightness conservation or the optical flow equation \citep{Neggers2016DIC}. To get rid of this spuriousness in the data, the entire dataset is first minmax-scaled (see Fig. \ref{fig:radio-aia}b), such that the scaled brightness of each pixel $i$ in the dataset is given by the following equation,
\begin{equation}
    B_{i,scaled} = \frac{B_i-B_{min}}{B_{max}-B_{min}}
    \label{eq:minmax}
\end{equation}
where, $B_{min}$ and $B_{max}$ are the minimum and maximum brightness of the entire dataset. This step ensures the entire dataset has brightness values $\in~[0,1]$, thereby standardizing the dataset. The resulting dataset is then subjected to the method of image correlation similar to as outlined in \cite{Mishra2024} and \cite{Routh2024} to obtain the sidereal rotation rate ($\Omega_{\theta}$), corresponding to the maximum correlation coefficient in each latitudinal bin. A brief description of the method is discussed in \autoref{sec:app}.

\section{Results and discussion}
\label{section:sec3}
\begin{figure}
    \centering
    \includegraphics[width=\linewidth]{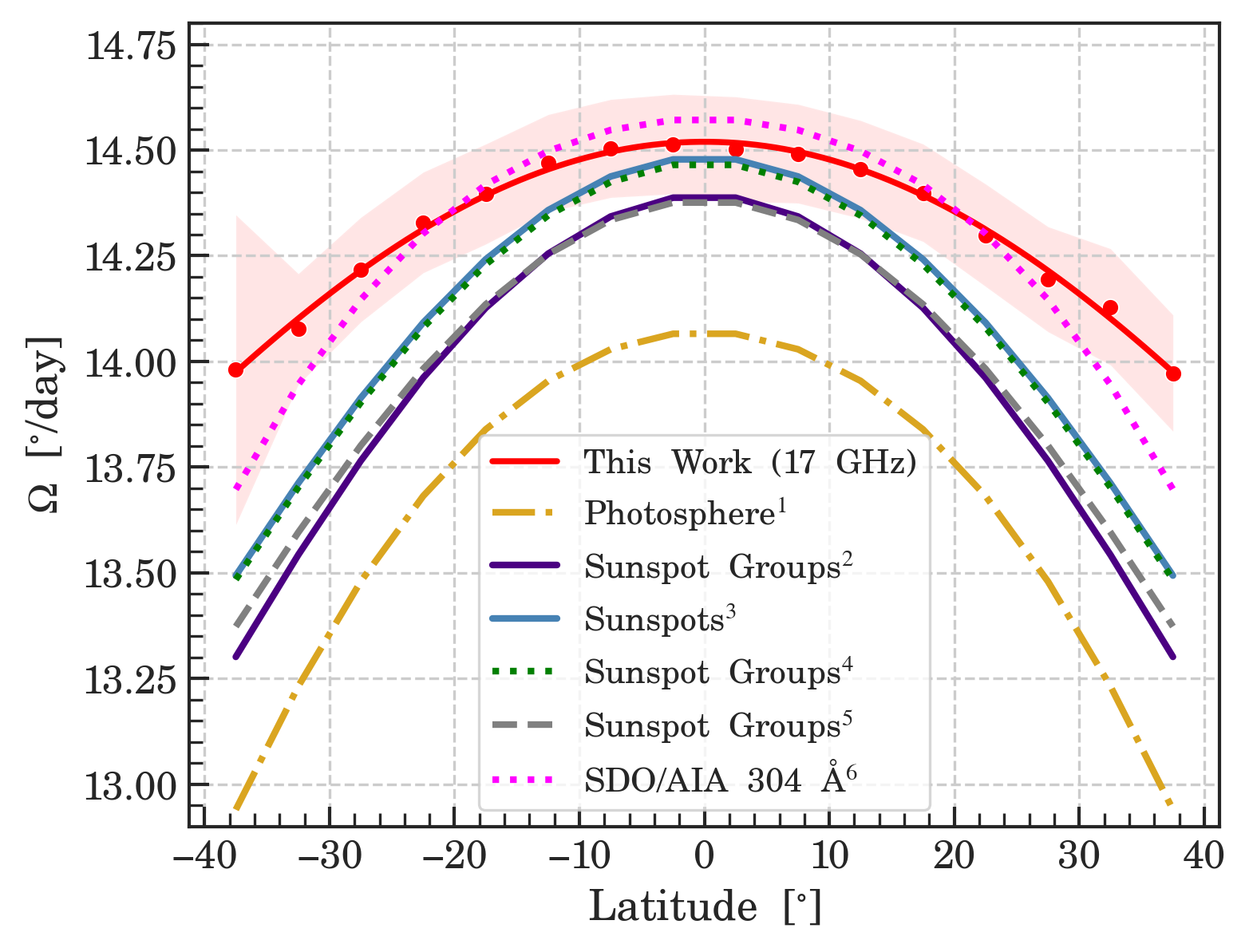}
    \caption{Rotation profile of solar chromosphere as obtained from 17 GHz data from Nobeyama Radioheliograph data, when compared with rotational profiles from $^{1}$\cite{Snodgrass1983,Snodgrass1984}, $^{2}$\cite{Howard1984}, $^{3}$\cite{beljan2017solar}, $^{4}$\cite{Ruzdjak2017}, $^{5}$\cite{Jha2021} and $^{6}$ \cite{Routh2024}.}
    \label{fig:rotprof}
\end{figure}

After setting a threshold on the value of cross-correlation coefficient (C.C.) to ensure the integrity of the analysis statistics (a detailed discussion of this approach and further examination of this technique is available in \cite{Mishra2024, Routh2024}), an average $\Omega_{\theta}$, weighted by the corresponding C.C. in each latitudinal bin is obtained for each latitude $\theta$. The uncertainty of these values is then determined as a combination of the standard statistical error ($\sigma_{SSE}$) and the least count error\footnote{$\sigma_{LCE}=0.1^{\circ}$}. These values are subsequently fitted to Eq. (\ref{diffequation}) using Levenberg-Marquardt least square \citep[LMLS;][]{Marquardt2009} method to obtain the best-fit parameters $A, B$ and $C$ and the respective uncertainty ($\Delta A, \Delta B$ and $\Delta C$) in determining them. Thereafter, we obtain the values $A+\Delta A = 14.520\pm0.006$ $^{\circ}/$day, $B+\Delta B = -1.443\pm0.099$ $^{\circ}/$day and $C+\Delta C = -0.433\pm0.267$ $^{\circ}/$day (Fig. \ref{fig:rotprof}) for the solar chromosphere as seen at a frequency of 17 GHz.  

\begin{figure}
    \centering
    \includegraphics[width=\linewidth]{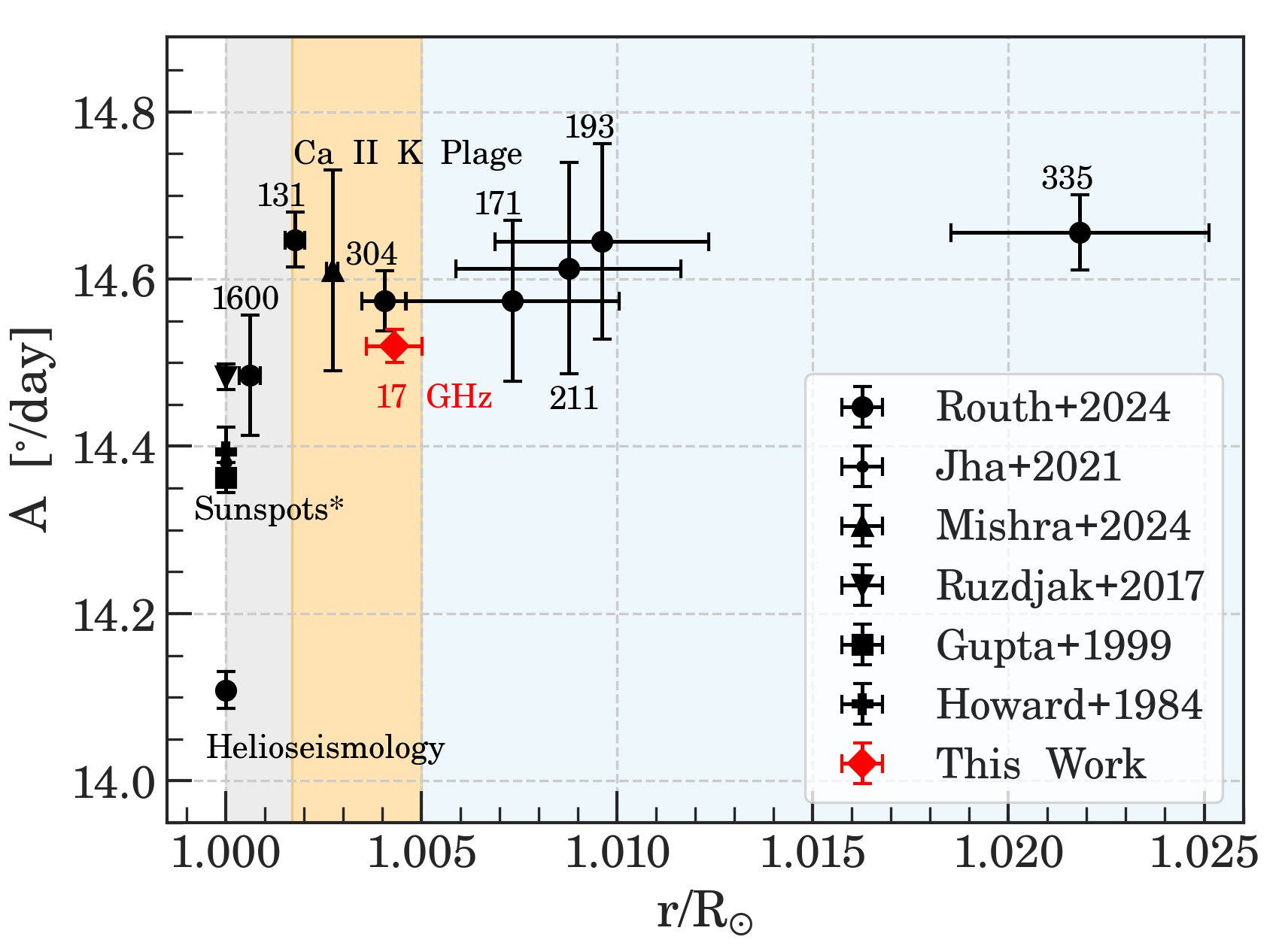}
    \caption{Variation in quatorial rotation rate from photosphere to different parts of corona from recent works. The umbrella label of "sunspots" is demarcated with an asterisk (*) to indicate different datasets used by different studies (see \autoref{app:comp}). Extents of different layers of the solar atmosphere are shaded differently (photosphere in gray, chromosphere in gold, and corona in light blue) to allow distinction between the same. The heights representing the different temperature sensitive filters have been taken from \cite{Sanjay2024} (for 131 {\AA}\, only) and the references in \cite{Routh2024}. The representative height for 17 GHz has been obtained from \cite{zirin1988astrophysics}. The errorbars along y represent 3$\sigma$ variation in the parameters, while along x the errorbars represent the exact variation reported in the mentioned studies. A detailed table compiling all values in this figure and several other studies are available for reference in \autoref{app:comp}.}
    \label{fig:heightwise}
\end{figure}

Upon examining Fig. \ref{fig:rotprof}, it is immediately apparent that the rotational profile is significantly higher than that of the photospheric plasma and sunspot groups, and it is relatively flatter, indicating that the solar chromosphere at a height of $3000 \pm 500$ km above the solar visible surface rotates more rapidly than the underlying photosphere at all latitudes, similar to findings by \cite{ChandraNobeyama2009} using three years of NoRH data. It is also evident that the rotational profile displays a relatively flatter trend, suggesting a reduced differential nature. This result aligns well with the rotational profile of the chromosphere from SDO/AIA $304$ {\AA} \citep{Routh2024}, which samples plasma in the temperature range of log$_{10}T=4.7$ K \citep{Lemen2012} and supposedly represents a height of $2820 \pm 400$ km \citep{Kwon2010} above the photosphere. Additionally, these findings are consistent with several studies such as \cite{Li2020,Mishra2024} that utilized spectroheliogram data from both space and ground-based data probing different parts of the solar chromosphere. The results for $A$ obtained in this study also conform to the existing trend for equatorial rotation rate at different heights in the solar atmosphere, as suggested by previous studies, as shown in Fig. \ref{fig:heightwise}. 

\begin{figure}
    \centering
    \includegraphics[width=\linewidth]{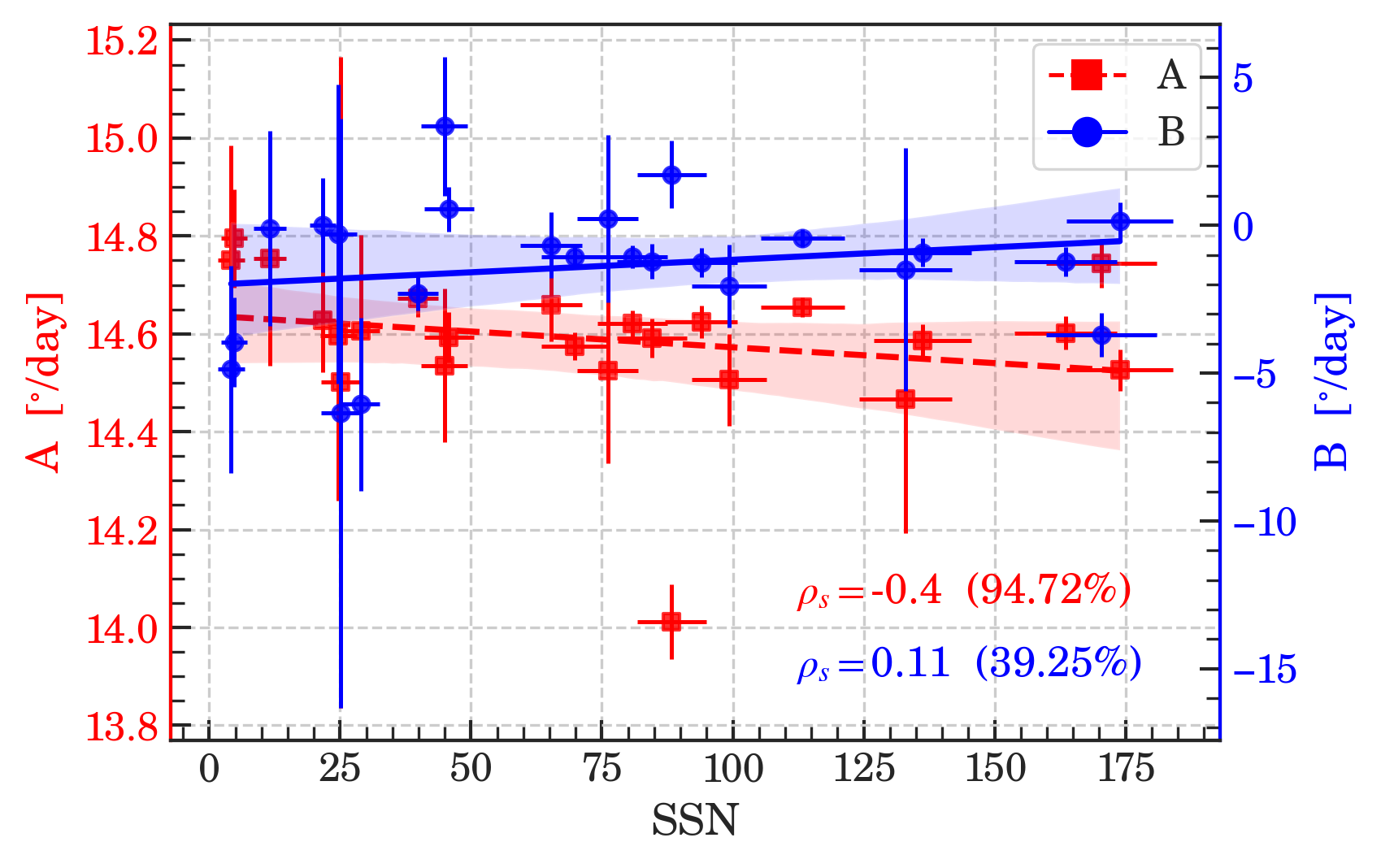}
    \caption{Correlation plot of equatorial rotation rate (A; in red) and latitudinal gradient (B; in blue) with yearly sunspot number and their error estimate in y and x directions, respectively. The lines of best fit are exhibited to visualize the generalized trend in the correlation. The shaded regions with each line correspond to the confidence interval of the fit.}
    \label{fig:yearly}
\end{figure}

Next, we look at the variation of the rotational parameters with respect to the strength of the solar cycle, as designated by sunspot number (SSN). From a correlation analyses involving the Spearman correlation coefficient ($\rho_{s}$) to explore any potential monotonous relationship, we find that both the equatorial rotation rate ($A$) and the latitudinal gradient ($B$) exhibit little to no significant relationship with solar activity (Fig. \ref{fig:yearly}). This observation is consistent with previous studies similar to \cite{Bertello2020,Mishra2024,Routh2024}. However, upon closer inspection, a very weak negative correlation for $A$ ($\rho_{s}=-0.4~(94.72\%)$) with solar activity strength is observed, aligning with \cite{Brajsa2006,Li2023}, but no such relationship is evident for $B$ ($\rho_{s}=-0.11~(39.25\%)$).

\section{Summary \& conclusions}
\label{section:sec4}
We analyzed 28 years of radio imaging data from the Nobeyama Radio Observatory to examine the differential rotation profile of the upper chromosphere using an automated image correlation technique\footnote{The generalized code is available for open source use at : \url{https://github.com/srinjana-routh/Image-Correlation}}. The distinguishing factor of this method is in the fact that unlike the tracer method which depends heavily on the availability of a singular feature distinguished by intensity, the method of image correlation is sensitive to only intensity gradients in each pixel and does not depend heavily on the availability of dominant features. This ensures that even during minima periods, a proper shift is identified based only on the intensities of each pixel in a bin. However, this also allows for the rotational profiles of several prominent features like coronal holes, filaments and active regions to coexist, without being differentiated. A numerical analysis to differentiate the contribution of these features may be attempted in the future, but remains out of scope for this particular study. Our goal in this study was to associate the derived rotational profile with a specific height in the upper solar atmosphere, reducing the ambiguity often encountered with EUV observations, which are typically sensitive to particular temperatures, leading to greater uncertainty regarding the representative height above the photosphere as higher atmospheric features of similar temperature range may also contribute to the filter-recorded intensity. Our findings indicate that the solar atmosphere at a height of $3000\pm500$~km rotates significantly faster than the photosphere below and shares similar rotational characteristics with $304$ {\AA} from SDO/AIA. This agreement of parameters obtained from a temperature sensitive channel which can include contributions from features at different heights reaching log$_{10}T = 4.7$ K locally, is within the 3$\sigma$ range (Fig. \ref{fig:heightwise}) with values derived from radio emissions, which predominantly originate from a single height \citep{Selhorst2008,Silvaheight2016}. Such an agreement reduces ambiguity in the possibility that a genuine upward trend in the solar atmosphere with increasing height. It is also worth noting that the overlap of the equatorial rotation rate ($A$) found in this study with that for $304$ {\AA}, which in turn matches the $A$ obtained for a depth of $r=0.94R_{\odot}$ \citep{Routh2024}, may further support the idea that magnetic features in the higher solar atmosphere have footpoints deep into the convection zone (e.g., \cite{Weber1969,Mancuso2020} and references therein). 

An investigation into the impact of solar activity strength on the parameters $A$ and $B$ revealed that neither parameter is significantly affected, although a very weak negative correlation was observed for $A$, in agreement with \cite{Li2013Solar-cycle-related,Jurdana2011,Ruzdjak2017, Wan2022Solar-cycle-related}.  This result may be understood in light of the numerical simulations explored by \cite{Brun2004, Brun2004ApJ}, whose results suggest that a subtle deceleration in differential rotation may be attributed to Maxwell stresses opposing Reynolds stresses, leading to reduced differential rotation. A similar result was obtained analytically by \cite{Lanza2006,Lanza2007} for young late-type stars. However, this inference should be taken with caution, as the correlation coefficient and its confidence level are too weak to draw any definitive conclusions.

This study's findings contribute to our understanding of the complex relationship between solar rotation and atmospheric height, by decreasing the ambiguity associated with the representative height associated with the observed rotational profile of higher solar atmosphere that has long been explored using optical, UV and EUV wavelengths. Further studies exploring coronal counterparts in wavelengths where height determination is less ambiguous could provide stronger confidence in the trend of rotation rate so consistently observed and thereby explore the mechanisms behind the observed rotational characteristics and their implications for solar dynamics. Additionally, long-term monitoring of these parameters could provide insights into potential changes in solar rotation over extended periods and as to why barely any impact of solar activity is seen on the chromospheric rotational rate, which may have further implications for solar activity cycles and space weather predictions.

\begin{acknowledgements}

S.R. is supported by funding from the Department of Science and Technology (DST), Government of India, through the Aryabhatta Research Institute of Observational Sciences (ARIES). The computational resources utilized in this study were provided by ARIES. AK acknowledges the ANRF Prime Minister Early Career Research Grant (PM ECRG) program. The data utilized in this work has been acquired from the Nobeyama Radio Observatory (NRO) database. All the authors are grateful to the observers at NRO for the data. Yearly mean sunspot numbers come from the source: WDC-
SILSO, Royal Observatory of Belgium, Brussels, and these can be
downloaded from \url{https://www.sidc.be/silso/}. 
\end{acknowledgements}

\begin{appendix}
\section{Segmentation and correlation analysis of the radio images}\label{sec:app}

\subsection{Method of image correlation}\label{subsec:appA}
\begin{figure}[h!]
    \centering
    \includegraphics[width=\linewidth]{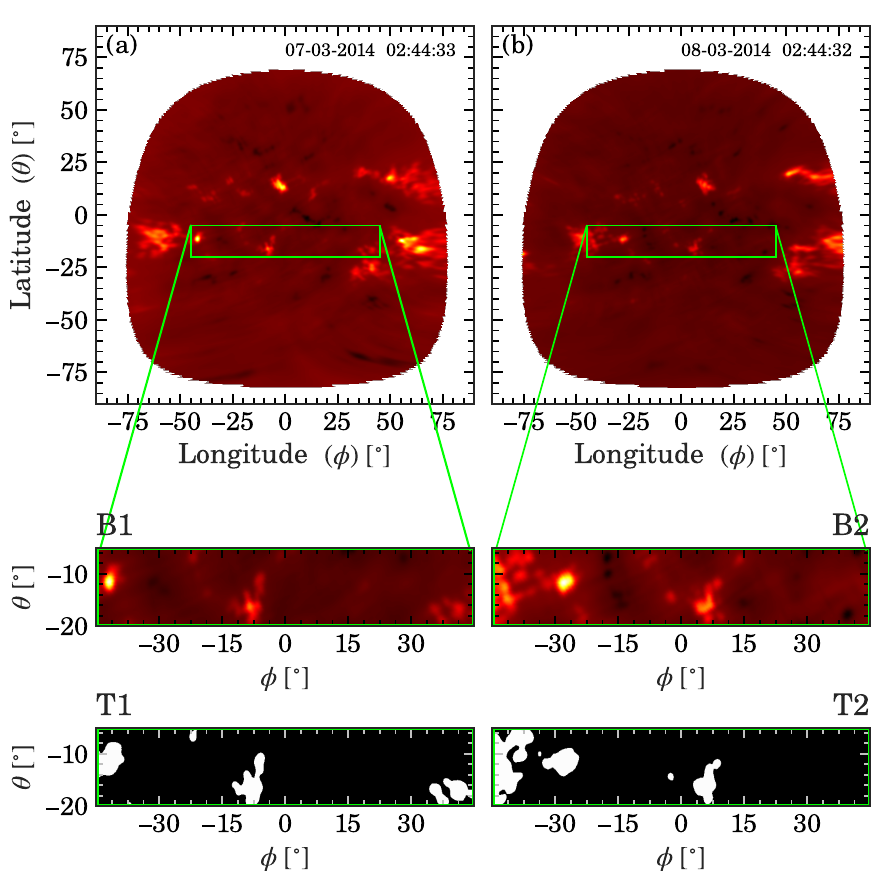}
    \caption{Set of images on (a) $7^{th}$ March, 2014  and (b) $8^{th}$ March, 2014 from the Nobeyama dataset after the conversion to Stonyhurst heliographic coordinates. Green rectangular boxes depict the bins B1 and B2, extending from $-20^{\circ}$ to $-5^{\circ}$ in latitude and $\pm 45^{\circ}$ in longitude each, on which image correlation is applied. The bins T1 and T2 depict the dominant bright features in the same bins that are majorly contributing to the correlation as demonstrated by adaptive intensity thresholding.}
    \label{fig:method_nobeyama}
\end{figure}

At a time a single pair of images, temporally separated by a day ($\Delta t \leq1$), are studied. These pair of images are first projected onto the heliographic grid $1800 \times 1800$ in dimension ($0.1^{\circ}$ along longitude and latitude; see Fig. \ref{fig:method_nobeyama}a and b).
The result is then divided in 16 overlapping bins of size $15^{\circ}$ and stride of $5^{\circ}$ in the latitudinal direction, spanning $\pm 45^{\circ}$ in latitude. Since there exists a center-to-limb variation in intensity which might affect our measurement of the longitudinal shift, the longitudinal span of these bins were restricted to $\pm 45^{\circ}$, thereby close to the disc centre where the effects of centre-to-limb variation is minimal (see for e.g., \cite{sudar2019centre} and references therein). At a time, bins of the same latitudinal extent temporally separated by 1 day (B1 and B2 in Fig. \ref{fig:method_nobeyama}) are subjected to the method of image correlation. In this method, one bin is shifted with respect to the other bin within the range of $\Delta\phi\in\Delta\phi_{0}\pm3^{\circ}$ and $\Delta\theta\in\Delta\theta_{0}\pm1^{\circ}$ in longitudinal and latitudinal directions, respectively\footnote{This analysis is performed using \href{https://hesperia.gsfc.nasa.gov/ssw/gen/idl_libs/astron/image/correl_images.pro}{correl\_images.pro} available in the SolarSoftWare library in Interactive Data Language (IDL)}. The initial guesses ($\Delta\theta_0, \Delta\phi_0$) were calculated from the rotation rate of sunspot groups \citep{Jha2021}. An analysis locating the maximum of the resulting 2D correlation coefficient gives the value of the longitudinal shift ($\Delta\phi$), as dictated by the dominant features, when present in the given latitudinal bin (see T1 and T2 in Fig. \ref{fig:method_nobeyama}; the method of segmentation of these features are discussed in \autoref{subsec:appA}). This value is then utilized to obtain the synodic value of $\Omega$ ($\Omega^{syn}_{\theta} = \frac{\Delta\phi}{\Delta t}$) at the mid-latitude $\theta_{m}$ of the same latitudinal bin. To incorporate the effect of the motion of the Earth around the Sun, we apply the sidereal correction \citep{rosa1995, wittman1996, skokik2014,Mishra2024} on the synodic rotation rate to get the sidereal rotation rate ($\Omega_{\theta}$ , which is used in the further analyses. A detailed discussion of this method is available in \cite{Mishra2024}. 

\subsection{Rotational profile estimates for different initial guesses.}
\begin{figure}][h!]
    \centering
    \includegraphics[width=\linewidth]{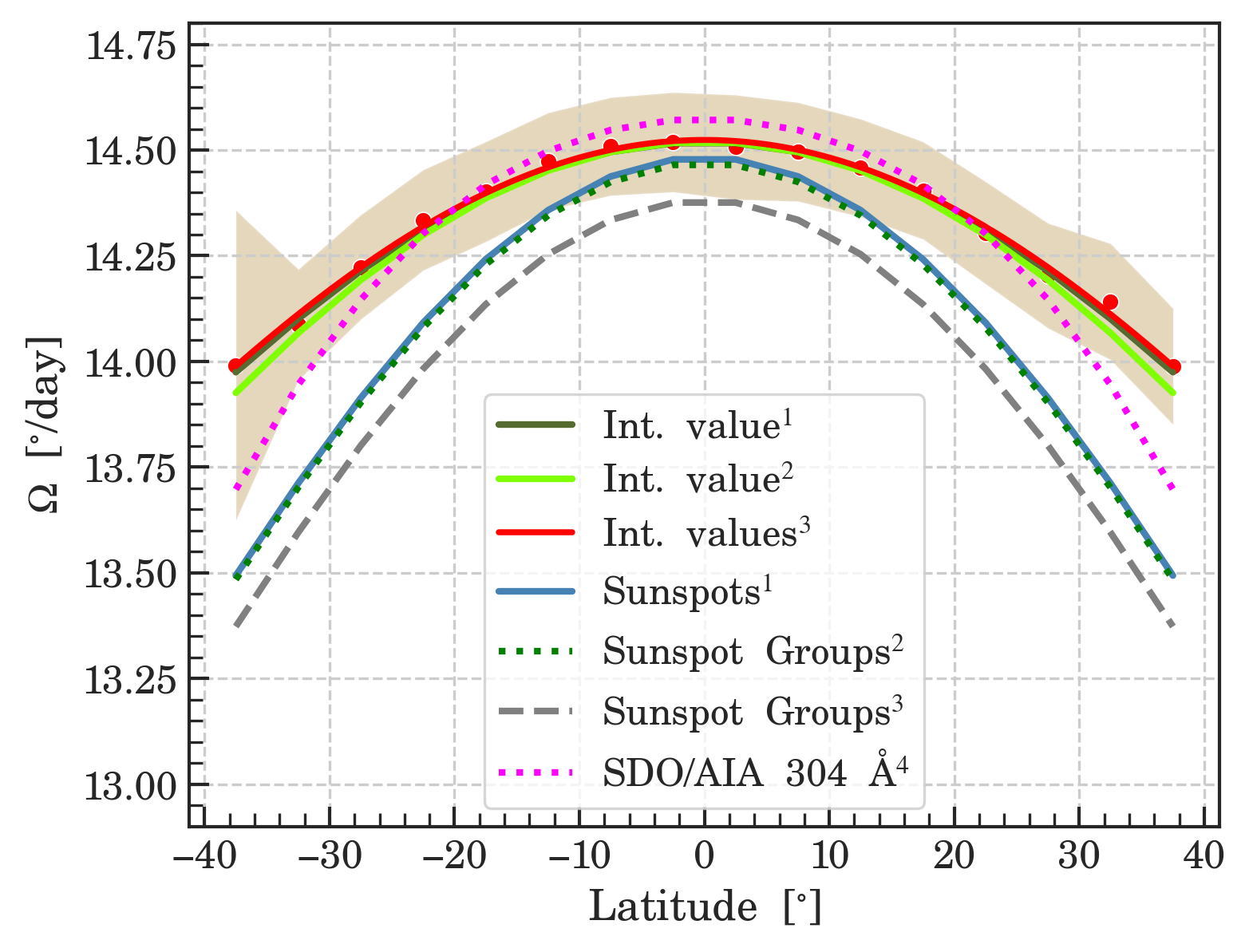}
    \caption{Obtained rotational profiles for Nobeyama 17 GHz data for different initial values from $^{1}$\cite{Ruzdjak2017} and $^{2}$\cite{beljan2017solar} along with the result from  the initial values taken from $^{3}$\cite{Jha2021}, as discussed in the main text, compared with the values from $^{4}$\cite{Routh2024}.The shaded region of each colour for initial guesses denotes the error pertained in the individual values for each case, as discussed in \autoref{section:sec3}.}
    \label{fig:new_initial_values}
\end{figure}

The entire analysis was re-run with different initial values from \cite{beljan2017solar, Ruzdjak2004} and compared with the original analysis, where initial values were taken from \cite{Jha2021} to see if the initial guesses contributed to any significant change in the rotational profiles thus obtained. As can been seen in Fig. \ref{fig:new_initial_values}, all the rotational profiles obtained overlap and there is no significant change even when the initial guesses obtained from the three depicted studies vary significantly (see \autoref{tab:int_val}).
\begin{table*}[h!]
    \centering
    \caption{Rotation parameters for entire span of Nobeyama 17 GHz data obtained using different initial guesses from different studies.}
    \begin{tabular}{cccc}
    \hline
    Initial Values from & $A\pm\Delta A$ & $B\pm\Delta B$ & $C\pm\Delta C$ \\
        & ($^{\circ}$/day) & ($^{\circ}$/day) & ($^{\circ}$/day) \\
    \hline
    \cite{beljan2017solar} & 14.523 $\pm$ 0.007 & -1.37 $\pm$ 0.11 & -0.20 $\pm$ 0.29\\
    \cite{Ruzdjak2017} & 14.519 $\pm$ 0.007 & -1.37 $\pm$ 0.11 & -0.25 $\pm$ 0.29\\
    \cite{Jha2021} & 14.520 $\pm$ 0.006 & -1.44 $\pm$ 0.10 & -0.43 $\pm$ 0.27\\[0.9ex]
    \hline
    \end{tabular}
    \label{tab:int_val}
\end{table*}
\subsection{Segmenting large-scale features}\label{subsec:appB}

To accentuate the understanding of the contribution of different features at different temperatures (or heights), an image segmentation method based on adaptive intensity thresholding was designed for the bright regions (e.g, active regions) (T1 and T2 in Fig. \ref{fig:method_nobeyama}). This was achieved by using the mean method for adaptive thresholding \citep{Chatterjee2016} to a smoothed and histogram-equalized image\footnote{The generalized code is available for open source use at \url{https://github.com/srinjana-routh/Bright-Regions-Nobeyama}} and is given by:
\begin{equation}
    I_{ij}=
    \begin{cases}
      1, & \text{if}\ I_{ij}\geq (\mu_{Disc}+~k\:\sigma_{Disc}) \\
      0, & \text{otherwise}
    \end{cases}
 \end{equation} 
 
where $I_{ij}$ is the intensity value of the $ij^{th}$ pixel, $\mu_{disc}$ is the mean, $\sigma_{disc}$ is the standard deviation of the solar disc and k ($=1$, here) is a scalar that can be adjusted based on the attribute we aim to segment. This method segments even small-scale brightenings which might correspond to coronal hole regions as well. An additional area threshold of $120.6$ arcsec\textsuperscript{2} ($\approx63.38 \times 10^{6}$ km\textsuperscript{2}), determined using trial-and-error method, was applied to discard any spurious brightening that might be of non-physical origin and extract large-scale features that remained for the exemplar latitudinal bin. 

\section{Rotation parameters of different parts and features of the solar atmosphere compiled from relevant studies}
\label{app:comp}

\begin{landscape}
\begin{table}[ht!]
    \centering
     \caption{Some rotation parameters of different features and parts of the solar atmosphere as obtained through various studies. (KSO: Kanzelhoe Observatory; KoSO: Kodaikanal Solar Observatory; MWO: Mt. Wilson Observatory; White Light: WL, Debrecen Photoheliographic Data: DPD, US Air Force Solar Optical Observing Network and National Oceanic and Atmospheric Administration:  USF/NOAA)}
    \begin{tabular}{ccccc}
    \hline
    Article & Data & $A\pm\Delta A$ & $B\pm\Delta B$ & $C\pm\Delta C$ \\
        &   & ($^{\circ}$/day) & ($^{\circ}$/day) & ($^{\circ}$/day) \\
    \hline\\[0.2ex]
    \multirow{2}{*}{\cite{Howard1984}} & Individual Spots (MWO; 1921-1982) & 14.522 $\pm$ 0.004 & -2.84 $\pm$ 0.04 & --- \\
    & Spot groups (MWO; 1921-1982) & 14.393 $\pm$ 0.010 & 2.95 $\pm$ 0.09 & --- \\[0.9ex]
    
    \cite{Balthasar1986} & Individual Spots (GPR; 1874-1976) & 14.551 $\pm$ 0.006 & -2.87 $\pm$ 0.06 & --- \\[0.9ex]
    
    \multirow{6}{*}{\cite{howard1999measurement}*} & Spot groups (Uncorrected KoSO data; 1917–1985) & 14.547 $\pm$ 0.005 & -2.96 $\pm$ 0.05 & --- \\
    & Spot groups (Uncorrected MWO data; 1917–1985)  & 14.459 $\pm$ 0.006 & -2.99 $\pm$ 0.06 & --- \\
    & Spot groups (Corrected KoSO; 1917-85) & 14.461 $\pm$ 0.006 & -3.02 $\pm$ 0.06 & --- \\
    & Spot groups (Corrected MWO; 1917-85) & 14.470 $\pm$ 0.005 & -2.97 $\pm$ 0.06 & --- \\
    & Individual Sunspots (Corrected MWO; 1917-85) & 14.446 $\pm$ 0.003 & -2.78 $\pm$ 0.03 & --- \\
    & Individual Sunspots (Corrected KoSO; 1917-85) & 14.456 $\pm$ 0.002 & -2.88 $\pm$ 0.02 & --- \\[0.9ex]
    
    \multirow{6}{*}{\cite{gupta1999}*} & Area < $5~\mu$Hem (KoSO; 1906-1987) & 14.491 $\pm$ 0.003 & -2.85 $\pm$ 0.03 & --- \\
    & 5 $\mu$Hem < Area < 15 $\mu$Hem (KoSO) & 14.380 $\pm$ 0.004 & -2.84 $\pm$ 0.04 & --- \\
    & Area > 15 $\mu$Hem (KoSO) & 14.279 $\pm$ 0.005 &  -2.83 $\pm$ 0.04 & --- \\
    & Area < 5 $\mu$Hem (MWO; 1917-1985) & 14.477 $\pm$ 0.003 & -2.80 $\pm$ 0.03 & --- \\
    & 5 $\mu$Hem < Area <15 $\mu$Hem (MWO) & 14.363 $\pm$ 0.006 & -2.65 $\pm$ 0.05 & --- \\ 
    & Area > 15 $\mu$Hem (MWO) & 14.248 $\pm$ 0.009 & -2.61 $\pm$ 0.09 & --- \\[0.9ex]
    
    \multirow{2}{*}{\cite{beljan2017solar}} & Spot groups (KSO sunspot drawings & 14.47 $\pm$ 0.01 & -2.66 $\pm$ 0.10 & --- \\
    & and WL images; 1964-2016) & 14.50 $\pm$ 0.01 & -2.87 $\pm$ 0.12 & --- \\[0.9ex]
    
    \multirow{5}{*}{\cite{Ruzdjak2017}*}& Individual Spots (GPR; 1874–1976) & 14.528 $\pm$ 0.006 & -2.77 $\pm$ 0.05 & --- \\
    & Individual Spots (USF/NOAA; 1977 – 2016 ) & 14.44 $\pm$ 0.01 & -2.54 $\pm$ 0.08 & --- \\ 
    & Individual Spots (DPD; 1977 – 2016) & 14.433 $\pm$ 0.009 & -2.44 $\pm$ 0.08 & --- \\
    & Individual Spots (GPR+USF/NOAA; 1874 – 2016) & 14.501 $\pm$ 0.005 & -2.71 $\pm$ 0.05 & --- \\
    & Individual Spots (GPR+DPD; 1874 – 2016) & 14.483 $\pm$ 0.005 & -2.67 $\pm$ 0.05 & --- \\[0.9ex]
    
    \cite{Jha2021}* & Binary masks of sunspots (KoSO; 1923–2011) & 14.381 $\pm$ 0.004 & -2.72 $\pm$ 0.04 & --- \\[0.9ex]
    
    \cite{Mishra2024} & Ca {\sc ii} K (KoSO;1907-2007) & 14.61 $\pm$ 0.04 & -2.18 $\pm$ 0.37 & -1.10 $\pm$ 0.61 \\[0.9ex]
    
    \multirow{7}{*}{\cite{Routh2024}} & 304 {\AA} (SDO/AIA; 2010-2023) & $14.574 \pm 0.012$ & -1.52 $\pm$ 0.12 & -2.29 $\pm$ 0.22 \\
    & 1600 {\AA}  & $14.485 \pm 0.024$ & -1.61 $\pm$ 0.24 & -2.68 $\pm$ 0.45 \\
    & 131 {\AA}  & $14.649 \pm 0.014$ & -1.33 $\pm$ 0.20 & -2.99 $\pm$ 0.52 \\
    & 171 {\AA} & $14.574 \pm 0.032$ & -1.36 $\pm$ 0.29 & -2.65 $\pm$ 0.46 \\
    & 193 {\AA}  & $14.645 \pm 0.039$ & -0.92 $\pm$ 0.34 & -2.70 $\pm$ 0.56 \\
    & 211 {\AA} & $14.613 \pm 0.042$ & -0.50 $\pm$ 0.37 & -3.31 $\pm$ 0.60 \\
    & 335 {\AA} & $14.656 \pm 0.015$ & -0.96 $\pm$ 0.22 & -2.75 $\pm$ 0.60 \\[0.9ex]
    \hline
    \vspace{0.5ex}
    *Denotes the studies depicted in Fig. \ref{fig:heightwise}.
    \end{tabular}
   
    \label{tab:solar_observations}
\end{table}
\end{landscape}

\end{appendix}

\end{document}